\documentclass[preprint]{aastex}

\slugcomment{to appear in April 2001 Astronomical Journal}

\shortauthors{Gizis et al.}
\shorttitle{2MASS L Dwarf Companion}

\begin{document}

\title{A 2MASS L Dwarf Companion to the Nearby K Dwarf GJ 1048}

\author{John E. Gizis\altaffilmark{1}, J. Davy Kirkpatrick}
\affil{Infrared Processing and Analysis Center, 100-22, 
  California Institute of Technology,
  Pasadena, CA 91125 \email{gizis@ipac.caltech.edu,davy@ipac.caltech.edu}}
\author{John C. Wilson}
\affil{Space Sciences, Cornell University, Ithaca, NY 14853
\email{jcw14@cornell.edu}}

\altaffiltext{1}{Visiting Astronomer, Kitt Peak National Observatory, 
National Optical Astronomy Observatories, which is operated by
the Association of Universities for Research in Astronomy, Inc. 
(AURA) under cooperative agreement with the National Science Foundation.}

\begin{abstract}
We report the discovery of a probable L1 companion
to the nearby K2 dwarf GJ 1048 using the
Two Micron All-Sky Survey (2MASS).  
This source, 2MASSI J0235599-233120 or GJ 1048B, has 2MASS
near-infrared colors and absolute magnitudes consistent with
an early L dwarf companion with a projected separation of
250 A.U.  The L1 spectral type is confirmed by
far-red optical and low-resolution IR spectroscopy.  
We present evidence that
GJ 1048 is a young ($\lesssim 1$Gyr) system, and that GJ 1048B may be
a high-mass brown dwarf below the hydrogen-burning limit.  
Additional studies of
the GJ 1048 system will help constrain the characteristics of 
L dwarfs as a function of age and mass.  
\end{abstract}

\keywords{binaries: general --- 
solar neighborhood --- stars: low-mass, brown dwarfs}

\section{Introduction \label{intro}}

Brown dwarfs are very-low-mass dwarfs which are
unable to maintain a stable luminosity by  
hydrogen fusion.  Analysis of radial velocity surveys 
has identified a ``brown dwarf desert'' ---
although stellar and planetary companions to FGKM main sequence stars
are relatively common,  less than 1\%  of such stars have companions in the
range 0.005 - 0.08 $M_\odot$ within 3 A.U.\citep{mazeh,mb00}.
Although brown dwarfs are rare as short-period companions
to more massive stars, they are quite common 
in the field.  The Deep Near-Infrared 
Survey (DENIS), Two Micron All-Sky 
Survey (2MASS), and Sloan Digital Sky Surveys (SDSS) 
have been used to  identify
numerous {\it isolated} brown dwarfs \citep{d99,k99,k00,s99,b99} belonging
to both the new cool L and yet cooler T (methane) spectral classes.
A first cut at modeling indicates that these discoveries imply that 
isolated substellar objects likely 
have a comparable space density to hydrogen-burning stars
\citep{reidmf}.   This discrepancy between the frequency of 
short-period brown dwarf secondaries and isolated field brown dwarfs
raises questions about star formation and binary formation.  
It is not yet known, however, if the ``brown dwarf desert''
extends to wide separations.  \citet{mazeh} have already noted
that short-period ($P<3000$ days) systems with G-dwarf primaries 
have a mass ratio
distribution that is flat or even favors high mass ratios,
while wider systems favor low mass (small mass ratio) systems.

2MASS offers the opportunity to address the frequency
of brown dwarf companions
for separations sufficiently large that the low-luminosity
companions can be resolved.  The feasibility of such a project
has already been demonstrated by the discovery of 
two M dwarf stellar companions to G dwarfs \citep{gl376b,g00b},
two L brown dwarf companions to G dwarfs \citep{k00},
and one T brown dwarf companion to a K/M/M triple system
\citep{gl570d}, as well as 
our recovery of the L brown dwarf companion to the M dwarf G196-3
\citep{g1963b}  --- all by searches aimed at
identifying {\it isolated} dwarfs.  A definitive measurement
of the stellar and brown dwarf companion frequency requires
careful statistical analysis accounting for the
ages, distances, and other characteristics of the systems
searched, but in the meantime analysis of any L dwarf
in a multiple system is important because measurements of 
the primary's trigonometric parallax, composition, and age
constrains the L dwarf's properties.  

We report the identification and confirmation of an L dwarf
companion to the nearby K2 dwarf GJ 1048 (HD 16270).  Our measurements
of the L dwarf are described in Section~\ref{data} and the 
age and masses of the GJ 1048 system are discussed in  
Section~\ref{discussion}.

\section{Data\label{data}}

We used the 2MASS Working Database to search for 
sources without optical counterparts 
within 60 arcseconds of a star in the
preliminary Third Catalog of Nearby Stars \citep{gj91}.  
We then examined
each image to confirm that the putative companion
was not an artifact.  We have not yet
followed up all the candidates identified by the
search, so no statistics can be computed.  However,
we noted that one candidate near the K2 dwarf GJ 1048
had J-K$_s = 1.354 \pm 0.140$ and K$_s = 12.315 \pm 0.077$ -- consistent
with an early L dwarf at the distance of GJ 1048.
This object is named 2MASSI J0235599-233120 based on
its position and inclusion in the 2MASS Second Incremental
Release Point Source Catalog
\citep{2mass2}, but henceforth we refer to
it as GJ 1048B and the primary, whose
catalog entry is 2MASSI 0236007-233116, as GJ 1048A.  
GJ 1048B is 3.75 arcseconds south and 11.34 arcseconds
west of the primary, implying a physical separation on the sky
of 250 A.U.  

\begin{figure}
\epsscale{0.8}
\plotone{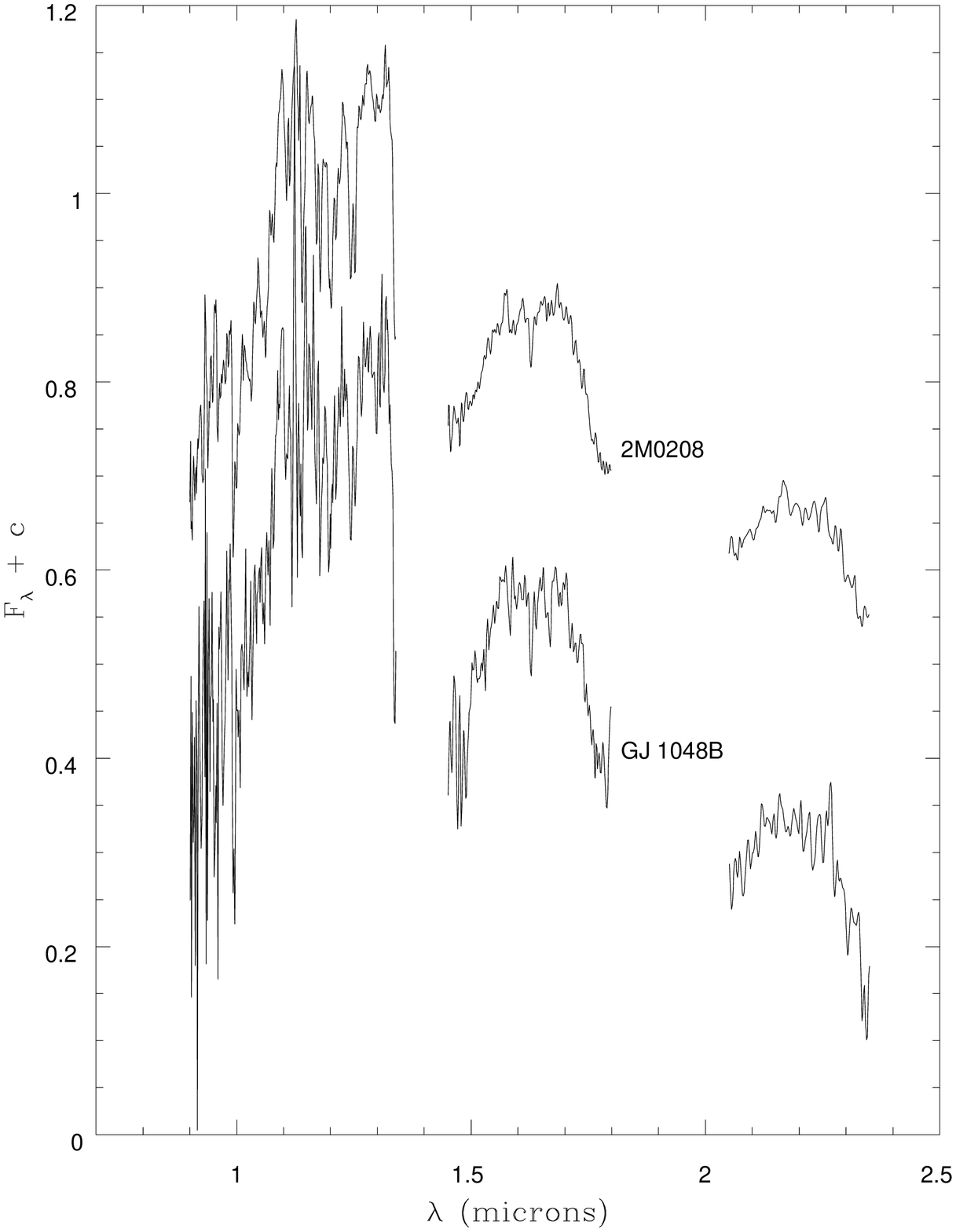}
\caption{CorMASS near-infrared spectra of GJ 1048B and 
2M0208, a known L1 dwarf.  2M0208 has been normalized to
agree with GJ 1048B and shifted upwards by 0.3.  
The relative fluxes between the J, H, and K band regions may
have small offsets.  
\label{fig-cormass}}
\end{figure}

\begin{figure}
\epsscale{0.8}
\plotone{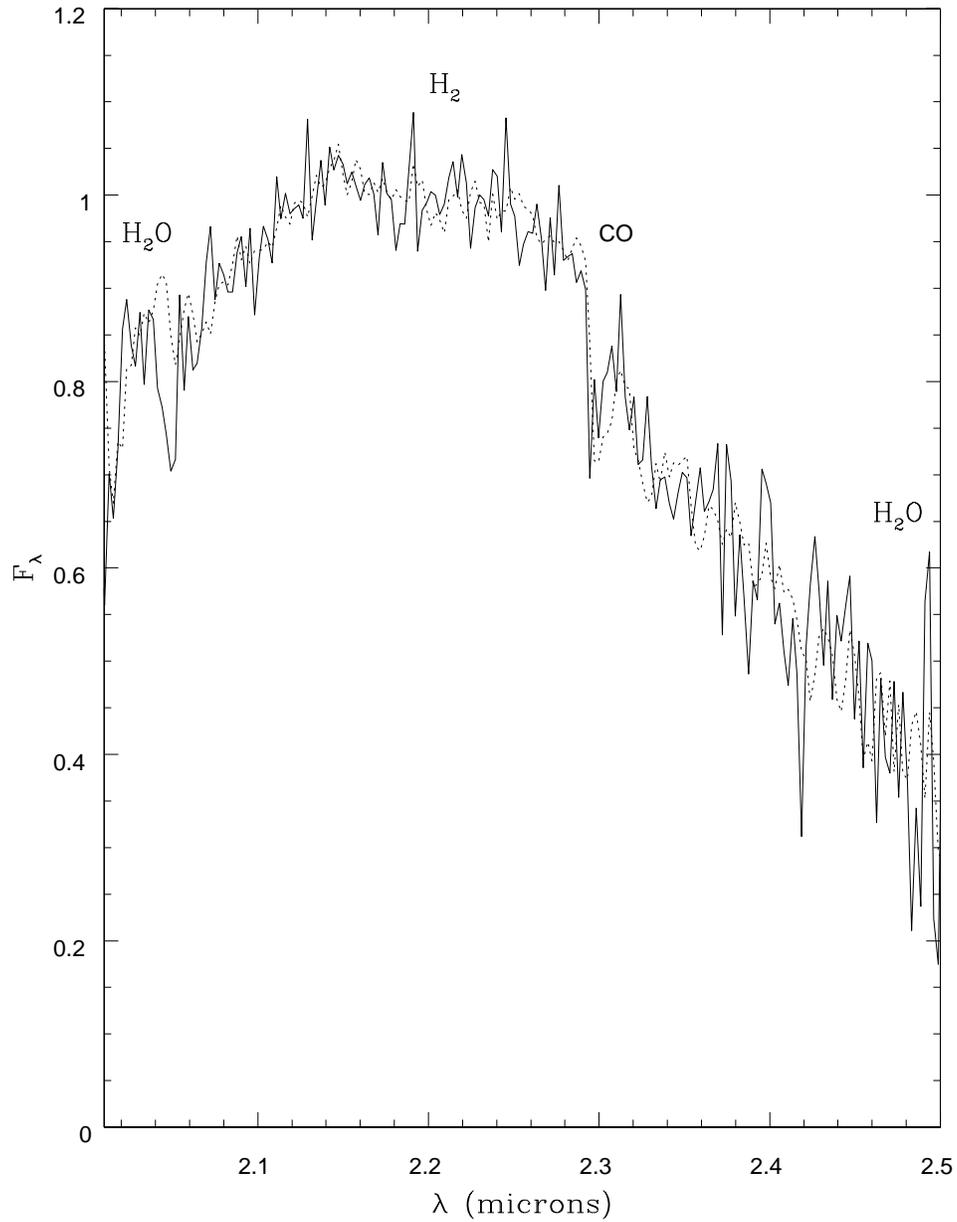}
\caption{Near-infrared K-band spectrum of GJ 1048B (solid line) compared to
the L2 dwarf 2MASSW J0015447+351603 (dotted line).  The two spectra
have been renormalized to unity over the range 2.14 -- 2.26 microns.  
\label{fig-kband}}
\end{figure}

We obtained low-resolution J and K band spectra 
on 27 November 1999 using the CRSP near-infrared
spectrograph on the Kitt Peak Mayall 4-meter telescope.
Details of the observing procedure and other L dwarfs
observed will be given in \citet{ldwarfir}.  
A low-resolution spectrum covering 0.9 to 2.4 microns was also 
obtained using the 
CorMASS infrared spectrograph \citep{cormass}
on the Palomar 60-inch telescope.  The
two spectra obtained with different 
instruments and reduction procedures are
in good agreement with each other.   Before obtaining
a far-red spectrum of GJ 1048B, a near-infrared spectral
type was estimated.  Available diagnostics include the slope of the J
band spectrum, the presence of FeH and \ion{K}{1} lines
in the J band \citep{nirspec}, and the strength of H$_2$O, H$_2$ and
CO absorption at K band \citep{tk99}.  Based on 
the comparison of these features to dwarfs
classified on the \citet{k99} system, an early L
spectral type was estimated.  
Fig.~\ref{fig-cormass} shows the CorMASS spectra
of GJ 1048B and the L1 dwarf 2MASSW J0208183+254253 \citep{k00}.  
Fig.~\ref{fig-kband}
compares CRSP spectra of GJ 1048B and the L2
dwarf 2MASSW J0015447+351603 \citep{k00}.

\begin{figure}
\epsscale{0.8}
\rotatebox{270}{
\plotone{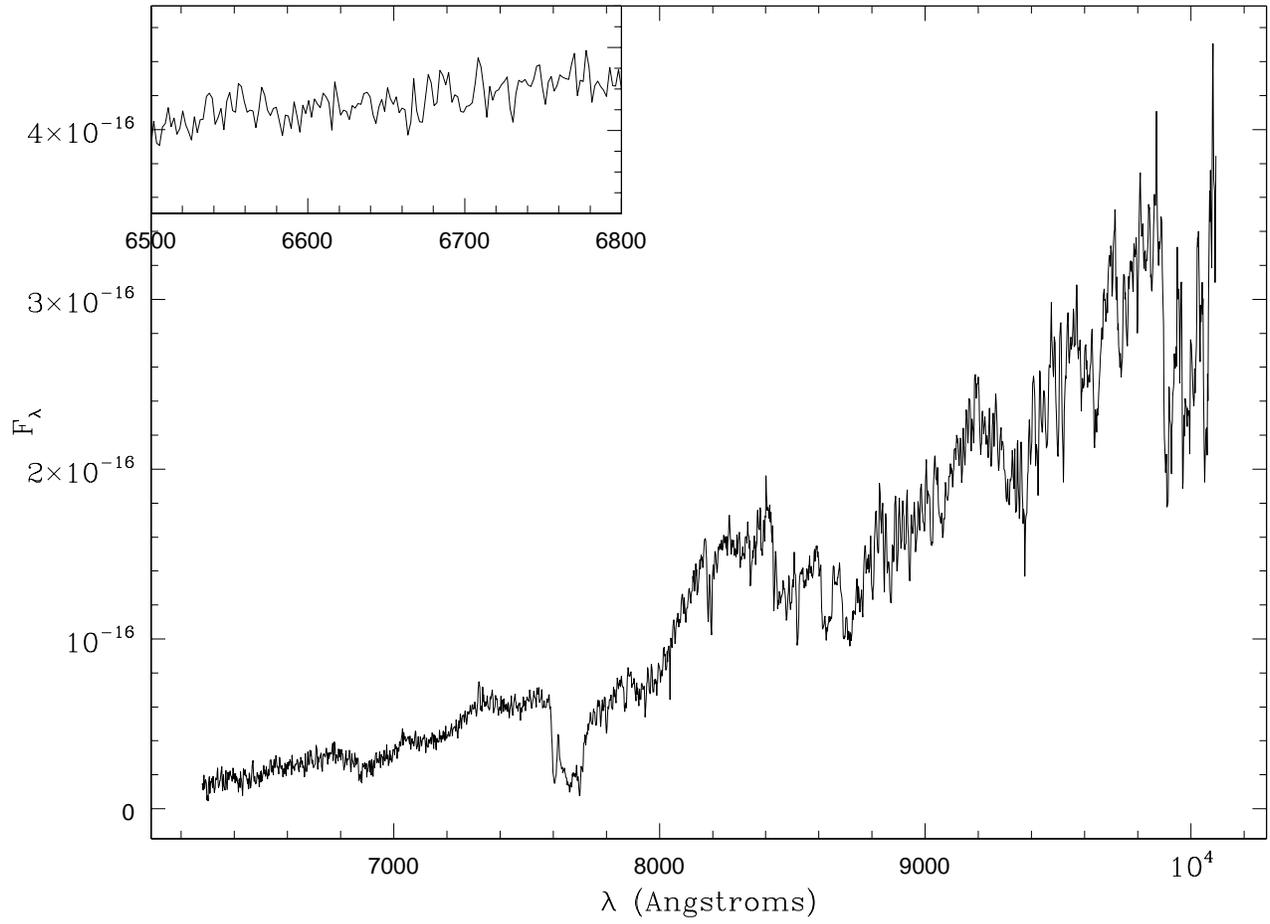}
}
\caption{Keck LRIS spectrum of GJ 1048B.  There is no evidence of
lithium absoprtion or H$\alpha$ emission.    
\label{fig-red}}
\end{figure}

On 23 August 2000 we obtained a Keck LRIS \citep{lris} 
spectrum of GJ 1048B (Fig.~\ref{fig-red}).  
The setup corresponded to that
used in \citet{k99}.  The \citet{k99} spectral type is
L1, and there is no detectable H$\alpha$ or lithium.  
The J-K$_s$ color is consistent with this spectral type
\citep[their Table 5]{k00}.

The accurate Hipparcos parallax of
$47.04 \pm 1.04$ milliarcseconds \citep{hipparcos}
implies $M_K = 10.677 \pm 0.09$.  \citet{k00}
have used the latest parallax data to derive an
L dwarf spectral type distance relationship.
For a spectral type L1, the relation gives $M_K = 10.60$
(the observed scatter in the relation 
is $\lesssim 0.5$ mags) which is in excellent agreement.  
The consistency of the absolute magnitude
with the color and spectral type supports our interpretation
that the L dwarf is a companion to the K dwarf.  

Given the uncertainties in 
the relation between spectral-type and temperature
\citep{k00,b00,l00}, GJ 1048B is likely to lie in
the range 1900-2200K.  Judging by the results of
\citet{tmr93}, \citet{jones94}, and
\citet{l00} for late-M and L dwarfs, a bolometric correction of 
BC$_K = 3.3 \pm 0.2$ is appropriate, implying 
$\log \frac{L}{L_\odot} = -3.7 \pm 0.1$.    

The proper motion of GJ 1048A is only 0.0855 arcseconds
per year, so confirmation that GJ 1048B has the 
same motion may be possible with careful observations 
within a few years but is not possible for 2MASS ---  it will
be four years until the system moves by $2\sigma$ (counting
only the first epoch uncertainty.) 
However, we can assess the probability that an unassociated
field L dwarf will be within 20 arcseconds of a cataloged
nearby star.  The surface density on the sky of 2MASS L dwarfs
to $K_s=14.7$ is 1 per 20 square degrees \citep{k99}, so 
the probability that a 2MASS L Dwarf lie within 15 arcseconds
any of the $\sim$4000 cataloged nearby stars is only 0.01.    
However, 80\% of the \citet{k99} L dwarfs lie at distances
greater than 30 parsecs, so the probability that an L dwarf
with a consistent distance estimate will be found is only  
$\sim 0.002$ over the entire sky.  It is therefore
highly likely that GJ 1048B is associated with the
nearby K dwarf.  

\section{Discussion\label{discussion}}

We may take advantage of our knowledge
of GJ 1048A to constrain the nature of GJ 1048B.  
GJ 1048A has estimated metallicity of $[Fe/H] \approx
-0.07$, a K2 dwarf spectral type, and $B-V = 1.07$
\citep{zs96,hipparcos}.  Using the observed V, J,
and K absolute magnitudes with mass-absolute magnitude
relationships based on binary star orbits\citep{hm93},
we find that GJ 1048A has a mass of $\sim 0.77$ M$_\odot$.

GJ 1048A is detected as a ROSAT X-ray source with
$\log L_X = 28.2$ \citep{rosatns}.  Since X-ray emission
is linked to age, we may estimate the age of the GJ 1048
system.  \citet{bmshr93} have analyzed the X-ray coronae
of field K dwarfs as a function of kinematic age.  GJ 1048A's
X-ray luminosity is typical for a ``Young Disk'' star  
and brighter than 95\% of ``Old Disk'' stars.  More
definite ages may be attached by using deep ROSAT 
pointings of open clusters.  GJ 1048A's corona is 
less luminous than all
dK stars in the Pleiades (age $\sim 10^8$ years) with
similar B-V colors \citep{scgph94}, so it is likely
older than that cluster.  GJ 1048A's corona is
comparable to the median X-ray luminosity observed
for single Hyades (age $\sim 6 \times 10^{8}$ years) K dwarfs 
\citep{pye}.  The uncertainty in this estimate is
necessarily large --- not only is there a range of X-ray luminosities
in any given cluster, but the similar age 
Praesepe and Ursa Major clusters show systematically
weaker X-ray emission than the Hyades \citep{rs95}.  
There are no available calibrating clusters near 1Gyr, so
we cannot set a definite upper limit to the age.  A factor
of 2 uncertainty seems appropriate.  
Our best estimate is that the GJ 1048 system has
an age of $\sim 6 \times 10^8$ years, though it is possible it
is as old as $\sim 10^9$ years.

We note that if GJ 1048A is actually a short-period binary,
its activity will be enhanced and it may be much older.  
According to the Simbad database, no radial velocity 
studies of GJ 1048A have been conducted.  The young age,
however, is supported by the kinematics of the
star.  The Hipparcos proper motion and parallax
\citep{hipparcos} 
implies $v_{tan} = 8.62$ km/s, which is characteristic
of a very young star.  

\begin{figure}
\epsscale{0.8}
\plotone{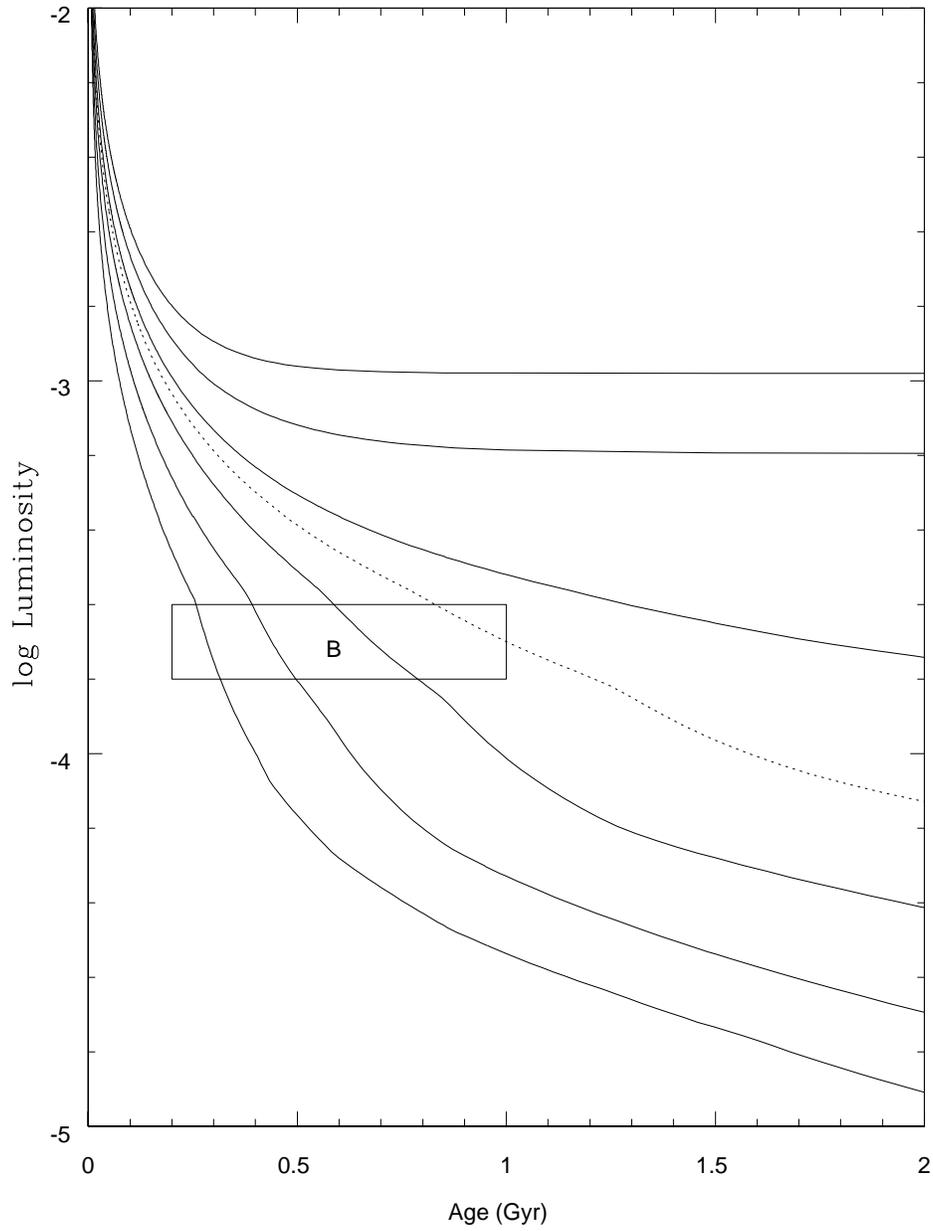}
\caption{\citet{burrows93,burrows97} models of the luminosity evolution
of stars and brown dwarfs.  From the left to right, the tracks 
represent 0.05, 0.06, 0.07, 0.076 (dotted), 0.08, 0.09,
and $0.10 M_\odot$ models.  The B marks our best estimate of
the age and luminosity of GJ 1048B, and the box shows our estimate
of the probable range of these parameters.
\label{fig-ev}}
\end{figure}

What does this age imply for the nature of GJ 1048B?  
Models that incorporate dust \citep{burrows93} indicate 
that the hydrogen-burning limit extends down to
masses of $0.076 M_\odot$, $T_{eff} \approx 1750$K and 
and $\log \frac{L}{L_\odot} \approx -4.2$.  
Without any age information, GJ 1048B is certainly consistent with being
a hydrogen-burning star.  
In the \citet{burrows93}, \citet{burrows97} and \citet{bcah98} models, however,
it requires $10^9$ years for 
a star at the hydrogen burning limit to cool to
the luminosity of GJ 1048B.  (Since the luminosity is
better determined than the temperature, we use it
for comparison to the models.)  Given our age estimates, it
is likely that GJ 1048B is a high-mass brown dwarf, but 
the data are also consistent with a star at the hydrogen-burning limit
at an age of 1 Gyr.  The younger
the age of the system, the more likely that GJ 1048B
is a brown dwarf (Fig.~\ref{fig-ev}).  The absence of lithium 
is not surprising, since  it only requires
$\sim 3 \times 10^8$ years for a $0.055 M_\odot$
model (the lithium-burning limit) to fade to the observed luminosity.   Given
the mass range $0.055-0.075$ M$_\odot$, the 
binary mass ratio is $q \approx 0.07 - 0.10$.  
The absence of H$\alpha$ emission in the L dwarf is also not
surprising since young L dwarfs tend to not show 
H$\alpha$ \citep{g00b}.  

Additional observations of this system are needed
to improve our knowledge of brown dwarfs.  Measurement
of the primary's radial velocity, rotation period,
rotation velocity, and chromospheric activity might 
improve the age estimate.  A search for a closer
companion, which could influence the activity 
level of the primary, is also needed. 

\section{Summary}

We have detected an probable L1 dwarf companion to the
K dwarf GJ 1048A.  Given the X-ray emission and low velocity 
of the primary, 
GJ 1048B is likely to be below the hydrogen-burning limit;
if a true star, it is within 0.005 M$_\odot$ of the limit.  
Studies of the primary may be able to further constrain
the age and status of the system.  As an object very near
the hydrogen-burning limit, GJ 1048B should serve as an
important object for further study.  

\acknowledgments

We thank the staffs of Kitt Peak and Palomar observatories
for their assistance in these observations, and the 
many people who have worked on making 2MASS a reality.  
CorMASS was made possible by
the work of Mike Strutskie, Jim Houck, J.D. Smith,
Michael Colonno,  Alan Enos, and Chuck Henderson.  
Adam Burrows, Gilles Chabrier, and Isabelle Baraffe
were kind enough to make their models available 
in electronic form.   JEG and JDK acknowledge the
support of the Jet Propulsion Laboratory, California
Institute of Technology, which is operated under contract
with NASA.  
This work was funded in part by NASA grant AST-9317456
and JPL contract 960847.
JCW acknowledges support by NASA grant NAG5-4376.
This publication makes use of data products from 2MASS, 
which is a joint project of the
University of Massachusetts and IPAC/Caltech, funded by NASA and NSF.
This research has made use of the Simbad database, operated at
CDS, Strasbourg, France.



\end{document}